\documentclass
[aps,pra,twocolumn,eqsecnum,superscriptaddress,showpacs,floatfix,nofootinbib]{revtex4}
\usepackage{graphicx}
\usepackage{bm}
\usepackage{dsfont}
\usepackage{amsmath,amsfonts,amssymb}
\usepackage{float}
\usepackage{hyperref}
\usepackage{amsmath}
\usepackage{amsfonts}
\usepackage{amssymb}\hypersetup{colorlinks=true,linkcolor=blue,citecolor=blue,urlcolor=blue}
\newcommand{\bi}{\begin{itemize}}
\newcommand{\ei}{\end{itemize}\normalsize}

\newcommand{\lr}[1]{\left( #1 \right)}
\newcommand{\ket}[1]{|#1\rangle}
\newcommand{\bra}[1]{\langle #1|}

\newcommand{\braket}[2]{\langle #1| #2 \rangle}

\newcommand{\no}{\nonumber}
\newcommand{\bea}{\begin{eqnarray}}
\newcommand{\eea}{\end{eqnarray}}
\newcommand{\be}{\begin{equation}}
\newcommand{\ee}{\end{equation}}

\newcommand{\stochmean}[1]{\langle\!\langle#1\rangle\!\rangle}
\begin{document}
\author{Hannes Pichler} \affiliation{Institute for Theoretical Physics,
  University of Innsbruck, A-6020 Innsbruck, Austria}\affiliation{Institute
  for Quantum Optics and Quantum Information of the Austrian Academy
  of Sciences, A-6020 Innsbruck, Austria}

\author{Johannes Schachenmayer} \affiliation{Institute for Theoretical Physics,
  University of Innsbruck, A-6020 Innsbruck, Austria}\affiliation{Institute
  for Quantum Optics and Quantum Information of the Austrian Academy
  of Sciences, A-6020 Innsbruck, Austria}
  \affiliation{Department of Physics and Astronomy, University of Pittsburgh, Pittsburgh, Pennsylvania 15260, USA}

\author{Andrew~J.~Daley} \affiliation{Department of Physics and Astronomy, University of Pittsburgh, Pittsburgh, Pennsylvania 15260, USA}

\author{Peter Zoller} \affiliation{Institute for Theoretical Physics,
  University of Innsbruck, A-6020 Innsbruck, Austria}\affiliation{Institute
  for Quantum Optics and Quantum Information of the Austrian Academy
  of Sciences, A-6020 Innsbruck, Austria}

\date{\today}
\title{Heating dynamics of bosonic atoms in a noisy optical lattice}

\begin{abstract}
We analyze the heating of interacting bosonic atoms in an optical lattice due
to intensity fluctuations of the lasers forming the lattice. We focus in
particular on fluctuations at low frequencies below the band gap frequency,
such that the dynamics is restricted to the lowest band. We derive stochastic
equations of motion, and analyze the effects on different many-body
states, characterizing heating processes in both strongly and weakly
interacting regimes. In the limit where the noise spectrum is flat at low
frequencies, we can derive an effective master equation describing the
dynamics. We compute heating rates and changes to characteristic correlation
functions both in the perturbation theory limit and using a full
time-dependent calculation of the stochastic many-body dynamics in one dimension based on
time-dependent density-matrix-renormalization-group methods.

\end{abstract}

\maketitle

\section{Introduction}

Ultracold atoms in optical lattices provide a clean and controllable
realization of quantum dynamics of an isolated many-body system on a lattice
\cite{Blo08,Lew_book,Jak05}. The remarkable progress in optical lattice physics
is underlined by recent experiments including the quantitative determination
of phase diagrams and critical phenomena of strongly interacting Hubbard
models \cite{Gre02,Hal09,Sch08,Jor10,Nas10,Zha12,Ku12,Sol12}, studies of quantum magnetism \cite{Sim11} and nonequilibrium quench dynamics \cite{Cir12,Blo12,Tro12}. A basic experimental challenge is the
preparation of low entropy or low temperatures states in optical lattices, and
to avoid possible heating mechanisms \cite{Jor10,Cir12,Blo12}. Heating can either be due to fundamental
decoherence sources like spontaneous emission \cite{Tro09,Pic10,Ger10,Pol12}, or collisional losses \cite{Gue06,Say08,Dal09}, and also
due to technical noise, for example, amplitude or phase noise of the lasers \cite{Lie03}
generating the optical lattice. While in a recent publication \cite{Pic12} we have
described possible optical lattice schemes which are immune to laser intensity
fluctuations, we will present below a detailed study of heating of bosonic
atoms in an optical lattice as a many-body non equilibrium problem.

We will study below heating within a \emph{single band Bose-Hubbard model}
where the tunneling and hopping parameters are stochastic functions of time
reflecting the intensity noise of the laser. We derive this model under the
assumption that the noise spectrum contains only significant components below
the band gap of the lattice, i.e.\ noise induced transitions to higher bands
can be neglected. In our model the intensity fluctuations of the light act as
a \emph{global noise}, which corresponds to the assumption that the spatial
correlations of the laser fluctuations are certainly much larger than the size
of the atomic cloud. The resulting stochastic schr\"{o}dinger Equation for the
Bose-Hubbard dynamics will be solved in detail in various limits and
approximations. First, we will compute the heating rates and the time
dependence of characteristic correlation functions in a perturbative
calculation valid for short times. In the white noise limit for the intensity
fluctuations we will be able to perform the stochastic average and derive a
master equation for the many-particle systems containing both the Hubbard
dynamics and the heating terms. In addition, we will solve the stochastic
many-body Schr\"{o}dinger equation in a Gutzwiller mean field approximation,
and in one dimension with a time-dependent density-matrix-renormalization-group (t-DMRG)
technique as a multiplicative stochastic differential equation. Besides
computing the total average energy transfer to the system as part of the
heating dynamics, we will also provide a detailed study of the excitations in
the many body system as signatures of the applied noise.

The paper is organized as follows. In Sec.~\ref{sec:Model} we derive a
stochastic Schr\"odinger equation for cold atoms in an optical lattice in the
presence of intensity noise. In Sec.~\ref{sec:Methods} we describe the methods
we use to analyze the resulting nonequilibrium dynamics, including full
time-dependent calculations based on t-DMRG calculations in one dimension (1D) and a
Gutzwiller mean-field approach for three-dimensional (3D) lattices. In Sec.~\ref{sec:Results} we
present the resulting time-dependent dynamics and discuss heating in different
parameter regimes, and in Sec.~\ref{sec:Summary} we present a summary and
outlook.

\section{Stochastic Many-Body Schr\"{o}dinger Equation}

\label{sec:Model} \begin{figure}[tb]
\includegraphics[width=0.46\textwidth]{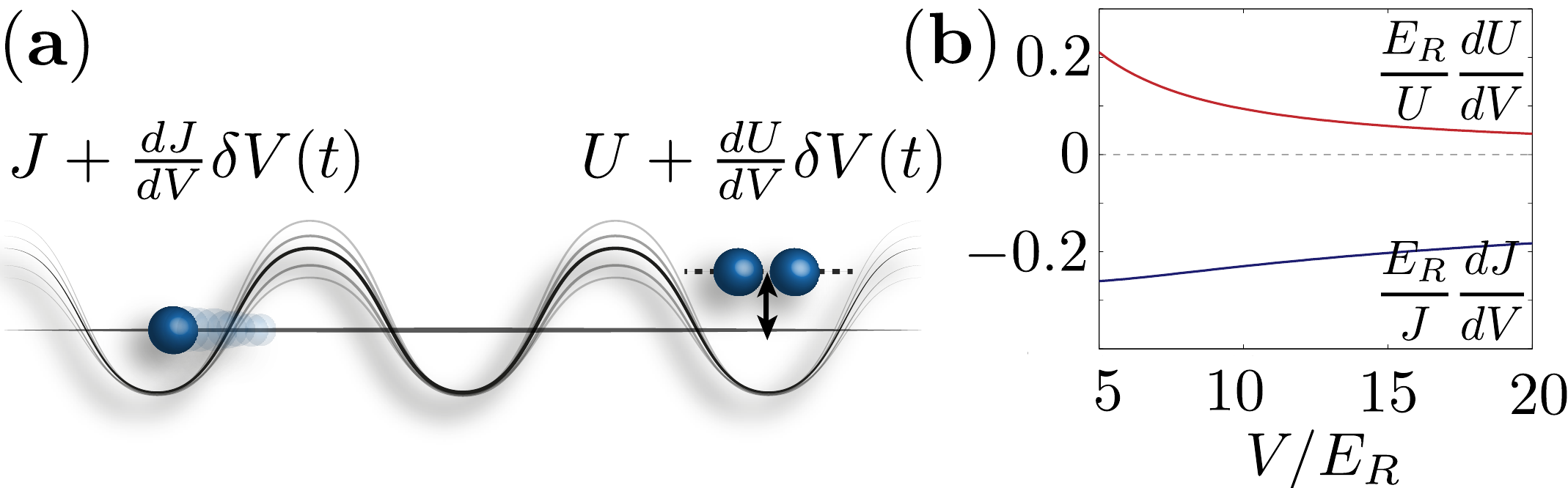} \caption{ (Color online) (a)
We consider bosons in an optical lattice where noise in the lattice depth
leads to noise in the hopping amplitude and in the interaction energy. (b)
Relative change of the hopping and the interaction parameter with lattice
depth in an isotropic three-dimensional cubic optical lattice, generated in the
standard way by three counter-propagating laser beams. Note that the change in
the hopping and interaction parameter is anticorrelated in this setup. }%
\label{Fig:1}%
\end{figure}

In this section, we derive and discuss a stochastic many-body Schr\"{o}dinger
equation (SMBSE) for ultracold bosonic atoms in an optical lattice in the presence of
intensity fluctuations of the laser generating the lattice. We are interested
in a situation where the atoms in the optical lattice are prepared in the
lowest Bloch band with dynamics described by a single band Bose-Hubbard
model \cite{Jak98},
\begin{align}\label{eq1}
H(J,U)=-J\sum_{\langle i,j\rangle}b_{i}^{\dag}b_{j}+\frac{U}{2}\sum_{i}%
b_{i}^{\dag}b_{i}^{\dag}b_{i}b_{i}.
\end{align}
Here the hopping amplitude and on-site interaction energy are denoted by
$J$ and $U$ respectively. The operators $b_{i}$ are the annihilation operators
for particles at site $i$. For deep lattices simple arguments give the dependence
$J\sim \frac{4}{\sqrt{\pi}}E_R\lr{V/E_R}^{3/4} \exp{(-2\sqrt{V/E_R})}$ and $U\sim 8 E_R (V/E_R)^{3/4}$ on the depth $V$ of the optical lattice (with $E_R$ being the lattice recoil energy). Thus an
increase of the lattice depth $V$ suppresses the tunneling, while at the same
time the on-site interaction becomes larger (Fig.~\ref{Fig:1}). The single-band tight-binding
model is valid provided the on-site interaction and temperature are much lower
than the gap to the first excited band. 

Laser intensity noise can be included in the Bose-Hubbard dynamics as
$V(t)=V_{0}+\delta V(t)$ with $\delta V(t)$ fluctuations around the mean
lattice depth $V_{0}$. For small fluctuations the tunneling  $J(t)=J_{0}%
+\frac{dJ}{dV}\delta V(t)$ and onsite interaction $U(t)=U_{0}+\delta V(t)\frac{dU}%
{dV}$ will become stochastic variables (see Fig.~\ref{Fig:1}), and we can
write a stochastic many-body Schr\"{o}dinger equation (SMBSE) ($\hbar=1$)
\begin{align}\label{eqSMBSE}
\frac{d}{dt}|\Psi\rangle &  =-i\left(  H(J_{0},U_{0})+H\left(  \frac{dJ}%
{dV},\frac{dU}{dV}\right)  \delta V(t)\right)  |\Psi\rangle\nonumber\\
&  \equiv-i\left(  H+H^{\prime}\delta V(t)\right)  |\Psi\rangle,
\end{align}
for a given noise model $\delta V(t)$. The derivatives $\frac{dJ}{dV}$, $\frac{dU}{dV}$ 
are evaluated at $V_0$ such that  $H\equiv H(J_0,U_0)$ and 
$H^{\prime}\equiv H(\frac{dJ}{dV},\frac{dU}{dV})$ are time independent. 
While $H$ induces coherent evolution according to the standard Bose-Hubbard 
Hamiltonian, $H'$ describes the heating
in the lowest lattice band due to intensity noise. We expect the above model to
be valid provided the fluctuation spectrum of $\delta V(t)$ is narrow on the
scale given the separation to the first Bloch band. Otherwise, the noise will
excite atoms to the higher Bloch bands. 

While the above heuristic derivation is intuitively obvious, we
summarize below in Sec.~\ref{sec:ModelA} a rigorous derivation of the above model starting
from a multiband Hubbard model, which establishes the validity of the SMBSE
given above, and gives corrections due to interband transitions. In Sec.~\ref{sec:White} we
will discuss the derivation of a master equation for the  averaged density
operator $\rho(t)=\stochmean{|\Psi(t)\rangle\langle\Psi(t)|}$ with
$\stochmean{\dots}$ denoting a stochastic average over the noise. This is
possible under the assumption of a white-noise limit, i.e., $\delta V(t)$ is
modeled by Gaussian white noise (within the single band model). In Sec.~\ref{sec:Methods}  we
will discuss mean-field and DMRG versions of the SMBSE, and their simulation.

Finally we note that similar discussions can be found in work on lattice spectroscopy \cite{Sto04,Iuc06,Kol06,Gre11,Tok11,Lac12}. There the potential is not fluctuating stochastically but modulated periodically in time. 

\subsection{Bose-Hubbard Model with Intensity Noise}

\label{sec:ModelA}  To derive Eq.~\eqref{eqSMBSE} We consider atoms (of mass $m$) in an
optical potential $V_{\mathrm{opt}}(x)=[V_{0}+\delta V(t)]\sin^{2}(kx)$, which
for simplicity of notation we assume to be one dimensional. Here $k$ is the wave
vector of the laser generating the lattice, which is related to the lattice
constant $a$ via $a=\pi/k$ and sets an energy scale via the recoil energy
$E_{R}=k^{2}/(2m)$. 
The full many body Hamiltonian, can be written in second
quantization using the bosonic field operators $\hat{\psi}(x)$ [$\hat{\psi
}^{\dag}(x)$] that destroy (create) a particle at the position $x$ as
\begin{align}
H(V(t)) &  =\int dx\,\hat{\psi}^{\dag}(x)\left(  -\frac{1}{2m}\frac{d^{2}%
}{dx^{2}}+V(t)\sin^{2}(kx)\right)  \hat{\psi}(x)\nonumber\label{eq2.6}\\
&  +\frac{g}{2}\int dx\,\hat{\psi}^{\dag}(x)\hat{\psi}^{\dag}(x)\hat{\psi
}(x)\hat{\psi}(x),
\end{align}
where $g=4\pi\hbar^{2}a_{s}/m$ and $a_{s}$ is the $s$-wave scattering length.

We are interested in the limit where the stochastic $\delta V(t)$ is much
slower than the (fast) time scale associated with the band gap. For a given
lattice depth $V$ the Wannier states $w_{j,n}(x,V)$ form a complete basis.
Thus it is natural to employ an adiabatic (Born-Oppenheimer) picture, where
field operators are expanded into \emph{instantaneous} Wannier states, that is
$\hat{\psi}(x)=\sum_{i,n}w_{j,n}(x,V(t))b_{i,n}(V(t))$, where $b_{i,n}(V(t))$
annihilates a boson in the instantaneous Wannier state $w_{j,n}(x,V(t))$ at
site $i$ in band $n$. In this way the single particle basis states keep track
of variations of $V(t)$ on a slow timescale. Non-adiabatic transitions due to
the time dependence of the basis states are driven by fast changes in the
lattice potential. As shown in Appendix \ref{app:Interband} the coefficients of the wave function $\ket{\Psi}$ in the
time-dependent Fock basis corresponding to this instantaneous Wannier states, $\langle\{n_{i,n}\}|\Psi\rangle$,
evolve according to
\[
i\frac{d}{dt}\langle\{n_{i,n}\}|\Psi\rangle=\langle\{n_{i,n}%
\}|H(V(t))+G(V(t))\dot{V}(t)|\Psi\rangle,
\]
with
\begin{align}
&  H(V)=-\sum_{n}J_{n}(V)\sum_{\langle i,j\rangle}b_{i,n}^{\dag}b_{j,n}%
+\sum_{n,i}\varepsilon_{n}(V)b_{i,n}^{\dag}b_{i,n}\nonumber\\
&  \qquad+\frac{1}{2}\sum_{i}\sum_{\{n\}}U_{\{n\}}(V)b_{i,n_{1}}^{\dag
}b_{i,n_{2}}^{\dag}b_{i,n_{3}}b_{i,n_{4}},\label{eq2.8}\\
&  G(V)=i\sum_{i,n;j,m}\int dx\,w_{j,m}(x,V)\frac{dw_{i,n}(x,V)}{dV}%
b_{i,n}^{\dag}b_{j,m}.\label{eq2.9}%
\end{align}
To simplify notation we suppressed the explicit dependence of the bosonic
operators and the Fock basis states on the instantaneous lattices depth
$V(t)$. An independent derivation and discussion of this equation in the context of a deterministically modulated lattice depth can be found in \cite{Lac12}.

Due to the localized nature of the Wannier functions only nearest-neighbor
hopping and on-site interactions are considered in~\eqref{eq2.8}. The first
term~\eqref{eq2.8} is simply the multi-band Bose-Hubbard
Hamiltonian with time-dependent parameters, corresponding to the
time-dependent lattice depth. The second term in~\eqref{eq2.9} arises from the
time dependence of the basis we use to describe the system. It is proportional
to the time derivative of the potential depth. In Appendix \ref{app:Interband} we give a
detailed discussion of this term and we show that it drives transitions between
different bands, but does not couple states within the same band. More
precisely, this \emph{interband} term mainly couples atoms to the second
excited band such that the number of atoms in the lowest band $N_{0}$
decreases in time as $\dot{N}_{0}/N_{0}\approx-2\eta^{4}S_{2}$. Here $S_{2}$
is the noise spectrum at the transition frequency from the lowest to the
second excited band and $\eta=\pi a_{0}/a\lesssim0.1$ is the Lamb-Dicke parameter
that compares the extension $a_{0}$ of the lowest band Wannier function to the
lattice constant $a$. This sets the timescale on which the restriction to the
lowest band is valid. For fluctuations that are slow on the time scale of
the gap this term is off resonant and can be dropped.

Under the same constraints also the first term in \eqref{eq2.8} can be restricted to the lowest
band and by linearizing the dependence of $J(V)$ and $U(V)$ for small
fluctuations on finds Eq. \eqref{eqSMBSE}. In the following we will describe the
heating dynamics on the basis of the single band model \eqref{eqSMBSE}.

\subsection{White Noise Approximation and Master Equation }\label{sec:White}

We are interested in calculating the response of the many-body system to the
noise stochastically averaged over the various realizations of 
$\delta V(t)$. This stochastic averaging can be performed exactly if we make
the assumption of white noise $\langle\!\langle\delta V(t)\delta V(t^{\prime
})\rangle\!\rangle=S_{0}\delta(t-t^{\prime})$. We note that this white noise
approximation implies that the fluctuations of $\delta V(t)$ are much faster
than $1/J$ and $1/U$, but are much slower than the transition frequency to the
first excited band. We can thus write the SMBSE as a Stratonovich stochastic
differential equation%
\begin{align}\label{eqStratonovich}
(S)\quad d|\Psi\rangle=-iH|\Psi\rangle dt-iH^{\prime}\sqrt{S_{0}}%
|\Psi\rangle d W_{t},
\end{align}

where $S_{0}$ denotes the strength of the noise and $dW_{t}$ is a Wiener
increment \cite{Gar_book,Klo_book}. This is a multiplicative differential equation, for which the
averaging over the noise can be performed exactly to derive a (master)
equation for $\rho=\langle\!\langle|\Psi\rangle\langle\Psi|\rangle\!\rangle$. 

To derive the master equation for $\rho$, we find it convenient to transform
the above equation to an Ito equation
\[
(I)\quad d|\Psi\rangle=\left(  -iH-\frac{S_{0}}{2}{H^{\prime}}^{2}\right)
|\Psi\rangle dt-iH^{\prime}\sqrt{S_{0}}|\Psi\rangle dW_{t}.
\]
Using the Ito rules for stochastic calculus \cite{Gar_book,Klo_book} one finds
the following many-body master equation:
\begin{align}\label{eqMaster}
\frac{d}{dt}\rho=-i[H,\rho]-\frac{S_{0}}{2}[H^{\prime},[H^{\prime},\rho]].
\end{align}
We note that this equation is of Lindblad form. The first term on the right
hand side is the familiar Bose-Hubbard Hamiltonian \eqref{eq1}, while the second term
describes heating. We note that the assumption of global intensity noise is
reflected in the \emph{spatially nonlocal} heating terms contained in the
double commutator. The above equation is derived from averaging over classical
noise (as opposed to coupling to a quantum reservoir). As a consequence
solutions $\rho(t)$ will in general approach for long times $\rho\sim\hat{1}$
corresponding to an (infinite temperature) completely mixed state (within the subspace allowed by the conserved quantities).

In Sec.~\ref{sec:Results} below we will derive analytical, perturbative solutions of the
master equation to describe the initial heating of a many-body quantum state
in the limit of weak ($U\ll J$) and strong interactions ($U\gg J$) . However,
for general parameters we find it more convenient instead of solving the
master equation numerically to compute averages from simulating trajectories
of SMBSE (c.f. Sec.~\ref{sec:Simulation}).

\subsection{Simulation of the Stochastic Many-Body Schr\"{o}dinger Equation}\label{sec:Simulation}

\label{sec:Methods} A simulation of the SMBSE as a multiplicative stochastic
differential equation can be performed in a mean-field limit and for 1D
systems using  t-DMRG techniques \cite{Vid03,Vid04,Dal04,Whi04} or exact state representation for small systems. 

The (mean field) Gutzwiller-ansatz \cite{Fis89,Rok91,Kra92} for the Bose-Hubbard model relies on a
product state assumption $|\Psi\rangle=\prod_{l}|\phi_{l}\rangle=\prod_{l}%
\sum_{n}f_{l,n}|n\rangle_{l}.$
The time-dependent variational ansatz \cite{Jak02,Zar05} for a homogeneous
system leads to a nonlinear Stratonovich stochastic Schr\"{o}dinger equation
of the form
\begin{align}
(S)\quad d|\phi_{l}\rangle=-iH_{l}|\phi_{l}\rangle dt-i\sqrt{S_{0}}%
H_{l}^{\prime}|\phi_{l}\rangle dW,\label{eq3.1}
\end{align}
where
\begin{align}
H_{l} &  =-zJ\left(  \psi_{l}^{\ast}b_{l}+\psi_{l}b_{l}^{\dag}\right)
+\frac{U}{2}{b_{l}^{\dag}}^{2}b_{l}^{2},\label{eq3.2}\\
H_{l}^{\prime} &  =-z\frac{dJ}{dV}\left(  \psi_{l}^{\ast}b_{l}+\psi_{l}%
b_{l}^{\dag}\right)  +\frac{1}{2}\frac{dU}{dV}{b_{l}^{\dag}}^{2}b_{l}%
^{2},\label{eq3.3}%
\end{align}
and $\psi_{l}=\langle\phi_{l}|b_{l}|\phi_{l}\rangle$. The number of nearest
neighbors is denoted by $z$.

To simulate the stochastic differential equation we typically used a
semi-implicit method given in \cite{Gar_book} of strong order 1.0. The
evolution in a small time step $\Delta t$ is calculated from
\begin{align}
&  |\bar{\Psi}_t\rangle=|\Psi_{t}\rangle-\frac{i}{2}H|\bar\Psi_{t}\rangle\Delta
t-\frac{i}{2}\sqrt{S}H^{\prime}|\bar\Psi_{t}\rangle\Delta W,\nonumber\\
&  |\Psi_{t+\Delta t}\rangle\approx2|\bar{\Psi}_t\rangle-|\Psi_{t}%
\rangle,\label{eq3.4}%
\end{align}
with a randomly chosen Wiener increment chosen from a normal distribution
$\Delta W=W(\Delta t)-W(0)\sim\mathcal{N}(0,\Delta t)$. The density matrix and
expectation values of operators are then obtained by averaging over the
trajectories calculated in this way. We directly incorporate this propagation
scheme for exact state representations and the analog version for the
Gutzwiller equations~\eqref{eq3.1}. 

Similar techniques can be employed in t-DMRG. There it is more convenient to
implement the propagation step in the form of a Trotter decomposition. Therefore,
we write the 1D Hamiltonian and noise term as sum over
next-neighbor operators $H=\sum_{i}H_{i,i+1}$ and $H^{\prime}=\sum
_{i}H_{i,i+1}^{\prime}$, respectively. For small time steps the evolution step
can then be implemented as
\begin{align}\label{eqTrotter}
|\Psi_{t+\Delta t}\rangle\approx\prod_{i}e^{-i\Delta tH_{i,i+1}}\prod
_{i}e^{-i\Delta WH_{i,i+1}^{\prime}}|\Psi_{t}\rangle.
\end{align}
To lowest order this is equivalent to the Euler algorithm and of weak order
1.0 convergence \cite{Klo_book,Gar_book}. Note that we did not find any
stability issues when using a number conserving update of the matrix product
states. We further confirmed that for sufficiently small timesteps the results
from the propagation \eqref{eqTrotter} coincide with the results obtained by exact
state representation from \eqref{eq3.4} for small systems ($N$ bosons on $M$
sites with $N=M\leq10$).

\section{Nonequilibrium many-body dynamics and heating}

\label{sec:Results}

In this section we show our results for the heating rates and analyze how the
noise changes the characteristics of the many-body ground state in the system.
We show how to obtain analytical results for heating rates in the two limiting
cases of a weakly interacting condensate in the superfluid (SF) phase ($U\ll J$) and of a nearly perfect
Mott insulator (MI) ($U\gg J$) and compare them to numerical simulations in
Sec.~\ref{sec:ResultsA}. In Sec.~\ref{sec:ResultsB} we analyze how the noise
changes the characteristics of the state and analyze the evolution of the
condensate fraction.

\subsection{Heating rates}
\begin{figure}[ptb]
\includegraphics[width=0.46\textwidth]{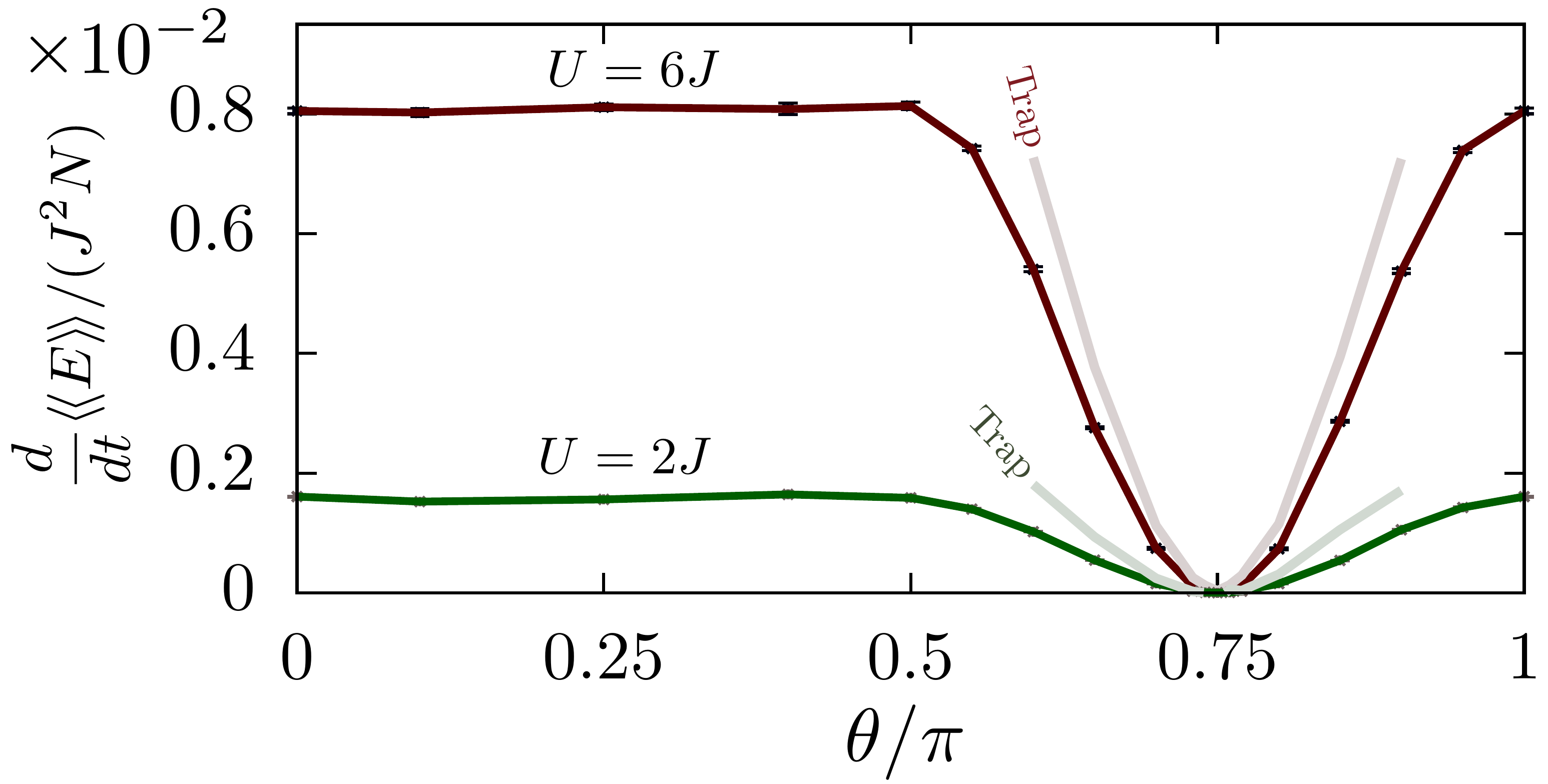} \caption{(Color
online) Short-time heating rates of superfluid ($U = 2J$) and Mott insulator
states ($U = 6J$) in one dimension as a function of the relative magnitude of noise on
$J$ and $U$. We parametrize the correlations between the noise on $J$ and $U$
by $\theta$ and $\lambda$ as $\sqrt{S}(dJ/dV)/J=- \lambda\cos^{2}(\theta)$ for
$0 \leq\theta< \pi/2 $; $\sqrt{S}(dU/dV)/U= \lambda\sin^{2}(\theta)$ and
$\sqrt{S}(dJ/dV)/J= \lambda\cos^{2}(\theta)$ for $\pi/2 \leq\theta< \pi$. The
usual anticorrelated case corresponds to $\theta<\pi/2$, and the sweet spot
$\xi=0$ of Eq.~\eqref{eq4.2} to $\theta=3\pi/4$. The heating rates are
calculated from linear regression over 500 t-DMRG trajectories in a system
with 30 particles on 30 sites (open boundary conditions). In both cases
heating is strongly suppressed in the vicinity of the sweet spot. Thin lines
show results in the presence of a harmonic trap with $\varepsilon_{i}/J =
0.0356 i^{2}$,$\sqrt{S_{0}}\frac{d\varepsilon_{i}}{dV}\frac{1}{\varepsilon
_{i}} =5\times10^{-3}J^{-1/2}$; in both cases we used $\lambda= 0.02J^{-1/2}$,
time step $\Delta t = 10^{-2}/J$. }%
\label{Fig:sweetspot}%
\end{figure}

If the density operator is diagonal in the eigenstates of the Hamiltonian
$H=H_{J}+H_{U}$ (where $H_{J}$ and $H_{U}$ denote the kinetic- and interaction-energy terms in the Bose-Hubbard model), for example if the system is in the ground or a thermal state, the average increase of the energy
$E=\langle H \rangle$ can be calculated from the master equation \eqref{eqMaster} and is
\begin{align}
\langle\!\langle\dot E\rangle\!\rangle & =\frac{S_{0}}{2}\left(  \frac{1}%
{J}\frac{dJ}{dV}-\frac{1}{U}\frac{dU}{dV} \right) ^{2}\langle\left[ \left[
H_{J},H_{U}\right] ,H_{J}\right]  \rangle. \label{eq4.1}%
\end{align}
The expectation value can be evaluated analytically in the limiting cases of
an ideal superfluid state, and an ideal Mott insulating state, as discussed below.

From~\eqref{eq4.1} we see that the heating vanishes if
\begin{align}
\frac{1}{J}\frac{dJ}{dV}=\frac{1}{U}\frac{dU}{dV}.\label{eq4.2}%
\end{align}
Only if this condition is met, the Hamiltonian $H$ and the noise operator
$H^{\prime}$ commute, moreover this means that they are proportional to each
other. As a consequence all states that commute with $H$, such as energy
eigenstates or thermal states, are stationary if~\eqref{eq4.2} is satisfied
(see Fig.~\ref{Fig:sweetspot}). In the standard setup, the hopping rate always
decrease with the lattice depth, while the onsite interaction always
increases, such that there is no such ``sweet spot'' [see Fig.~\ref{Fig:1}(b)]. However, one can come up with more elaborate lattice setups (see, for
example, \cite{Pic12}) that are designed in such a way to fulfill the sweet
spot condition and therefore are resilient against this type of noise. For
later convenience we introduce the parameter
\begin{align}
\xi=\left(  \frac{1}{U}\frac{dU}{dV}-\frac{1}{J}\frac{dJ}{dV} \right) ,
\end{align}
that measures the deviation from this sweet spot. 

\label{sec:ResultsA} \begin{figure}[tb]
\includegraphics[width=0.46\textwidth]{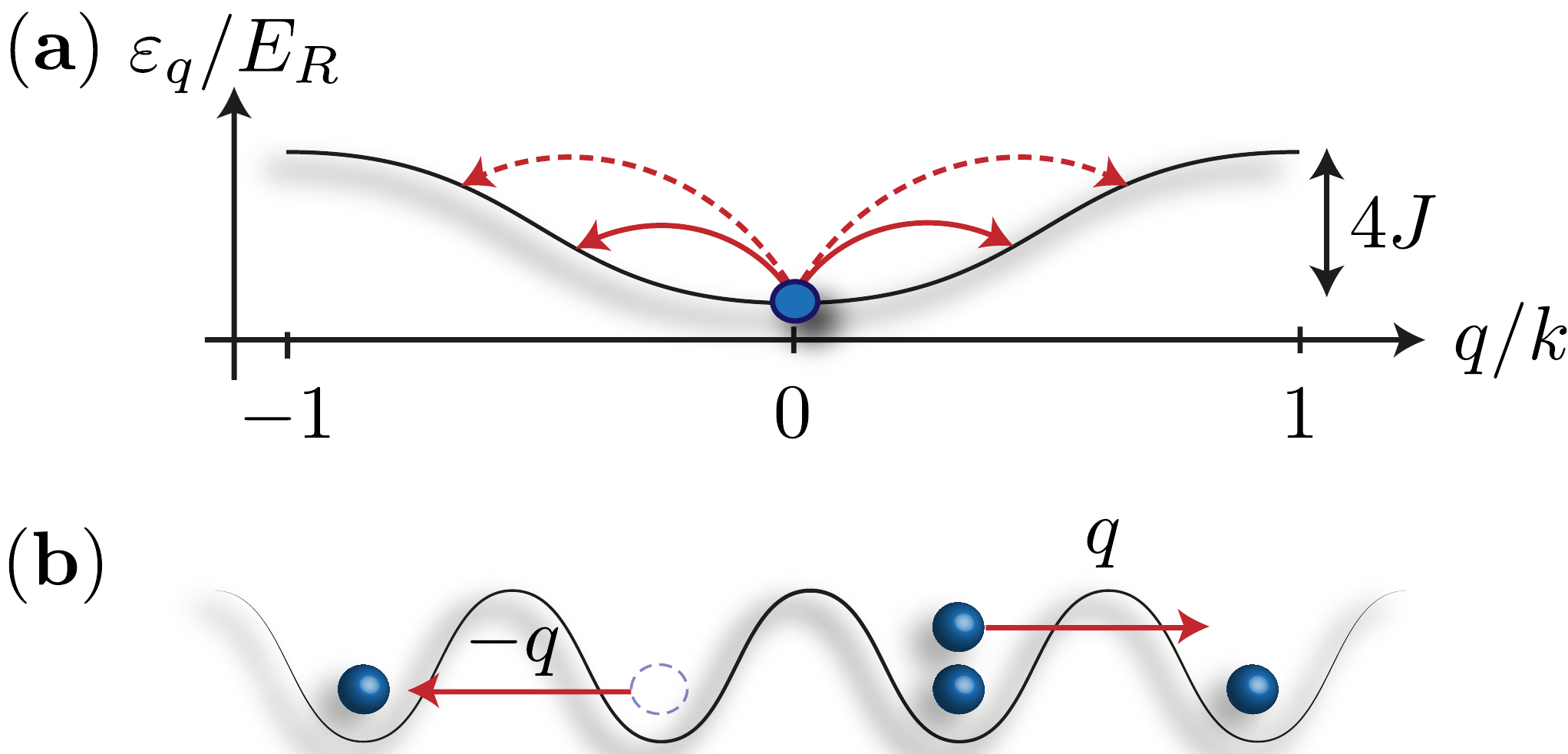}
\caption{(color online) Elementary heating processes in the limiting cases of
weak and strong interaction. (a) For weak interactions ($U\ll J$) lattice
fluctuations create pairs of Bogoliubov excitations with opposite
quasimomenta on top of the Bose-Einstein condensate at $q=0$. (b) In the
strongly interacting case ($U\gg J$) lattice fluctuations create pairs of
excess particles and holes of opposite quasimomenta $\pm q$ on top of the
Mott insulator. }%
\label{Fig:heating_processes}%
\end{figure}

\begin{figure}[tb]
\includegraphics[width=0.46\textwidth]{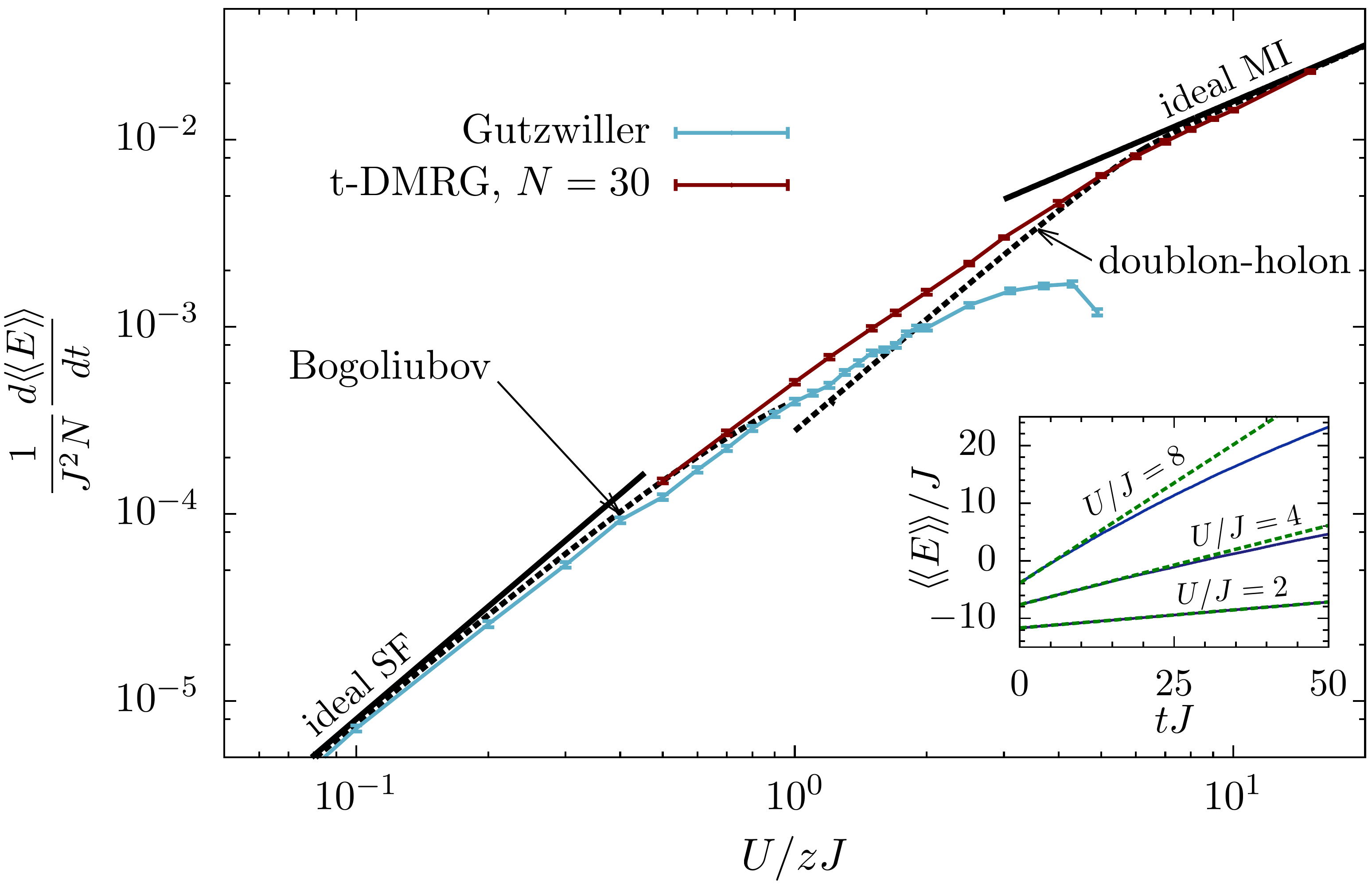}
\caption{(Color online) Comparison of the short-time heating rates from the
Gutzwiller ansatz and t-DMRG simulations to the analytical results for weak
and strong interactions. The Gutzwiller calculation is for a homogeneous
infinite system, the DMRG simulation for a 1D system of $30$ particles on $30$
sites. We consider a small anticorrelated noise ($\sqrt{S} \frac{1}{J}
\frac{dJ}{dV} = -0.01 J^{-1/2}$, $\sqrt{S} \frac{1}{U} \frac{dU}{dV} = 0.01
J^{-1/2}$). We average over $n_{t}=1000$ ($n_{t}=500$ for t-DMRG) noise
trajectories and estimate statistical errors of the mean. Convergence has been
checked with time-steps $1 \times10^{-3} \leq\Delta tJ \leq1 \times10^{-2}$
and Hilbert space truncations $8\leq d_{l} \leq16$. The t-DMRG results are
converged with a bond dimension of $D=200$. 
The inset shows the averaged mean energy as a function of time (solid lines) together with the linear increase (dashed lines) with heating rate given by eq.~\eqref{eq4.1}. This shows that the energy increase is well captured by a constant heating rate for several hopping times. The parameters and methods used to 
calculate the inset are the same as in Fig.~\ref{fig:cf}(b). }%
\label{Fig:heating_rates_comparison}%
\end{figure}

\subsubsection{Weak interactions}

If the interactions are weak, the ground state of the system is a
Bose-Einstein condensate, where a macroscopic number of atoms occupies the
mode with zero quasimomentum. For an ideal condensate 
we obtain from Eq.~\eqref{eq4.1} (for a cubic lattice with $z$ nearest neighbors and a filling of
$\bar n$ atoms per site) to lowest order in $U/J$ the heating rate per
particle:
\begin{align}
\frac{\langle\!\langle\dot E\rangle\!\rangle}{N} & =S_{0}\left(  \frac{1}%
{J}\frac{dJ}{dV}-\frac{1}{U}\frac{dU}{dV} \right) ^{2} z JU^{2}\bar
{n}.\label{eq4.3}%
\end{align}
To find corrections to this result we can apply a Bogoliubov approximation.
This approximation is most conveniently expressed in the Bloch basis rather
than in the Wannier basis, such that the kinetic part of the Hamiltonian reads
$H_{J}=\sum_{q}\varepsilon_{q} b_{q}^{\dag} b_{q}$, where the single-particle energy spectrum $\varepsilon_q$ 
for a cubic lattice in $d$ dimensions is $\varepsilon_q=2J\sum_{i=1}^{d}[1-\cos(q_i a)]$. 
Here $q_i$ denotes the component of the quasi momentum along direction $i$. The interaction is treated
on a mean field level, replacing
$H_{U}\rightarrow U\bar{n}/2\sum_{q}(2b_{q}^{\dag} b_{q}+b_{q} b_{-q}%
+b_{q}^{\dag} b_{-q}^{\dag})$.
The total Hamiltonian is then quadratic and can be diagonalized by a standard
Bogoliubov transformation $b_{q}=u_{q} c_{q}+v_{-q}c_{-q}^{\dag}$ such that
$H=\sum_{q}\tilde\varepsilon_{q} c_{q}^{\dag}c_{q}$, with $\tilde
\varepsilon_{q}=\sqrt{\varepsilon_{q}(\varepsilon_{q} + 2U\bar{n})}$.
Within the same approximation the noise operator $H^{\prime}=\frac{1}{J}%
\frac{dJ}{dV}H_{J}+\frac{1}{U}\frac{dU}{dV}H_{U}$ is also quadratic. However,
the Bogoliubov transformation diagonalizing $H$ does not diagonalize
$H^{\prime}$ [except if condition~\eqref{eq4.2} is met]. Therefore, the noise
operator contains terms $c_{q}^{\dag}c_{-q}^{\dag}$ and thus excites pairs of
Bogoliubov excitations with opposite quasimomenta
[Fig.~\ref{Fig:heating_processes}(a)]. The total quasimomentum is
conserved as required by the conserved symmetry. The heating rate associated
with this process can easily be calculated from~\eqref{eq4.1} for the vacuum
of quasiparticles. For a $d$-dimensional cubic lattice ($z=2d$) it is given
by
\begin{align}
\label{eq4.4}\frac{\langle\!\langle\dot{E}\rangle\!\rangle}{N} & =S_{0}\left(
\frac{1}{J}\frac{dJ}{dV}-\frac{1}{U}\frac{dU}{dV} \right) ^{2}2JU^{2}\bar
{n}G\left(  \bar{n}U/J \right) ,
\end{align}
with
\begin{align}
\label{eq4.5}G(\epsilon) & =\frac{1}{\pi^{d}}\!\!\!\!\int\limits_{0<x_{i}<\pi}
\!\!\!\!d^{d}x \frac{\left(  \sum_{i=1}^{d}[1-\cos(x_{i})] \right) ^{3/2}%
}{\left(  \sum_{i=1}^{d}[1-\cos(x_{i})]+\epsilon\right) ^{1/2}}.
\end{align}
In one dimension this integral can be calculated exactly and leads to
\begin{align}
\label{eq4.6}G(\epsilon) & =1-\frac{\epsilon}{2}+\frac{4}{\pi}\sqrt
{\frac{\epsilon}{2}}+\frac{1}{\pi}(2-\epsilon)\arcsin\left(  \frac{2-\epsilon
}{2+\epsilon} \right).
\end{align}
For arbitrary dimensions we can expand the integral and obtain (to first order
in $\epsilon$) $G(\epsilon)=d-\frac{\epsilon}{2}$, such that for
$U/J\rightarrow0$ the rates~\eqref{eq4.3} and~\eqref{eq4.4} coincide.

\subsubsection{Strong interactions}

According to Eq.~\eqref{eq4.1}, the heating rate for a perfect Mott insulator with an integer
number of atoms per site $\bar n$ is (to lowest order in $J/U$) given by  
\begin{align}
\label{eq4.7}\frac{\langle\!\langle\dot E\rangle\!\rangle}{N} & =S_{0}\left(
\frac{1}{J}\frac{dJ}{dV}-\frac{1}{U}\frac{dU}{dV} \right) ^{2} zUJ^{2}\left(
\bar{n}+1 \right) .
\end{align}
Also here we can obtain corrections as well as insight into the excitation
process by suitable approximations. Following \cite{Che12} one can
approximately describe the one-dimensional Bose Hubbard model in the Mott
regime by restricting the local Hilbert space to the three states $|\bar
{n}\rangle$ and $|\bar{n}\pm1\rangle$. Using a generalized Jordan-Wigner
transformation one introduces fermionic creation operators for the excess
particles ($c_{j,+}^{\dag}$) and holes $(c_{j,-}^{\dag})$. Assuming that the
density of excess particles and holes is small, the Hamiltonian $H=H_{J}%
+H_{U}$ and the noise operator $H^{\prime}=\frac{1}{J}\frac{dJ}{dV}H_{J}%
+\frac{1}{U}\frac{dU}{dV}H_{U}$ can be written in the quasimomentum basis
approximately as
\begin{align}
H_{J}  & \approx-2J\sum_{p} \cos(pa)[(\bar{n}+1)c_{p,+}^{\dag}c_{p,+}+\bar
{n}c_{p,-}^{\dag}c_{p,-}]\nonumber\\
& +\sqrt{\bar{n}(\bar{n}+1)}i\sin(pa)(c_{p,+}^{\dag}c_{-p,-}^{\dag}%
-c_{-p,-}c_{p,+}), \nonumber\\
H_{U} & \approx U\sum_{k} \bar{n}c_{p,+}^{\dag}c_{p,+}-(\bar{n}-1)c_{p,-}%
^{\dag}c_{p,-}.\label{eq4.9}%
\end{align}
The Hamiltonian $H$ is then quadratic and can be diagonalized by the
Bogoliubov transformation $\gamma_{\sigma p,\sigma}^{\dag}=u_{p} c_{
p,\sigma}^{\dag}+v_{p} c_{\sigma p,\sigma}$,
such that $H= \sum_{p,\sigma=\pm}\, \sigma\epsilon_{\sigma}(p)\gamma
_{p,\sigma}^{\dag}\gamma_{p,\sigma} $ with the quasiparticle dispersion
relation
\begin{align}
& \epsilon_{\pm}(p) = - J\cos(pa)+\frac{U}{2}(2\bar{n}-1)\nonumber\\
& \pm\frac{1}{2}\sqrt{[U-2(2\bar{n}+1)J\cos(pa)]^{2} + 16\bar{n}(\bar{n}+1)
J^{2} \sin^{2}(pa)}.\label{eq4.10}%
\end{align}
As in the weakly interacting case, the noise operator $H^{\prime}$ is in
general not diagonal in the quasiparticle basis, but it contains terms
$\gamma_{p,+}^{\dag}\gamma_{-p,-}^{\dag}$ that generate correlated
quasiparticle pairs that travel though the system with opposite
quasimomentum. To lowest order in $J/U$, these are pairs of excess atoms and
holes [Fig.~\ref{Fig:heating_processes}(b)]. The heating rate associated with
these processes can easily be calculated from~\eqref{eq4.1} with~\eqref{eq4.9}
and~\eqref{eq4.10} for the vacuum of quasiparticles. It is given by
\begin{align}
\label{eq4.11}\frac{\langle\!\langle\dot E\rangle\!\rangle}{N} & =S_{0}\left(
\frac{1}{J}\frac{dJ}{dV}-\frac{1}{U}\frac{dU}{dV} \right) ^{2} 2UJ^{2}\left(
\bar{n}+1 \right) F(J/U,\bar{n}),
\end{align}
where we abbreviated
\begin{align}
& F(j,n)=\frac{1}{\pi}\int_{0}^{\pi}dp \frac{2\sin^{2}(p)}{\sqrt{f(j,n,p)}%
},\label{eq4.12}\\
& f(j,n,p)=\left(  1-(4n+2)j\cos(p) \right)  ^{2}\!\!+16(n^{2}+n)j^{2}\sin
^{2}(p).\nonumber
\end{align}
For $j\ll1$ one finds $F(j,n)  =1+(1 - 2 n - 2 n^{2})j^{2}+\mathcal{O}(j^{4}%
)$. For $J/U\rightarrow0$ this reduces to~\eqref{eq4.7}.

\begin{figure*}[tbh]
\includegraphics[width=0.9\textwidth]{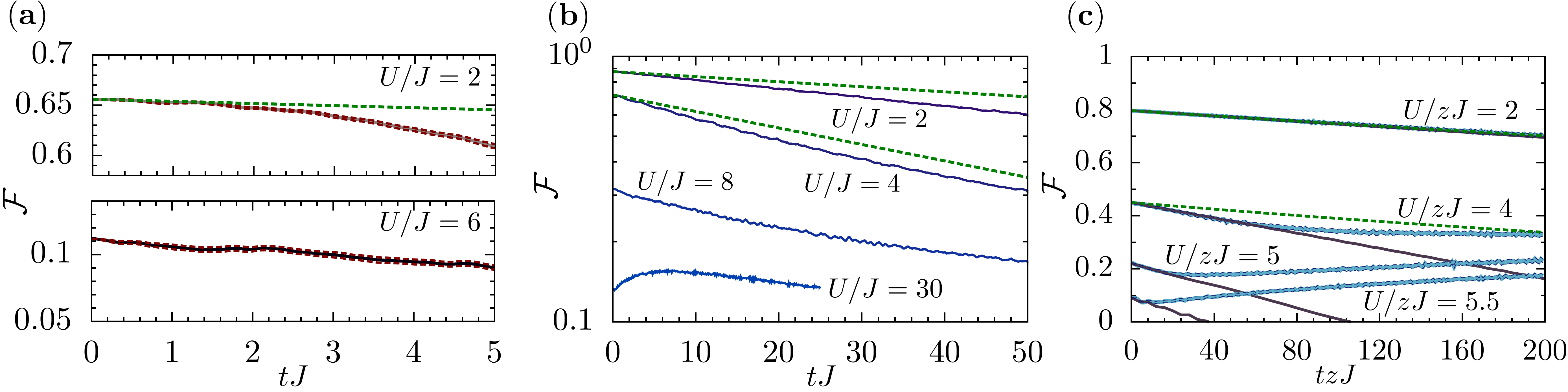}  \caption{(Color
online) (a) The time evolution of the condensate fraction for short times in a
large system with $M=N=48$ (averaged over $60$ trajectories, bond dimension
$D=256$, local dimension truncation $d_{l}=6$), (b) for long times in a small
system with $N=M=10$ ($500$ trajectories), and (c) for a homogeneous infinite
Bose-Hubbard model in mean-field theory (Gutzwiller ansatz, $d_{l}=8$, $1000$
trajectories, solid grey lines obtained by a small noise approximation
\cite{Gar_book}). The fraction in general decreases except for the extreme
Mott insulating case of $U/J=30$ and in the long time limit in mean-field
theory. In (a) the noise is anticorrelated with $\sqrt{S} \frac{1}{J}
\frac{dJ}{dV} = -0.025 J^{-1/2}$, $\sqrt{S} \frac{1}{U} \frac{dU}{dV} = 0.025
J^{-1/2}$, in (c) with $\sqrt{S} \frac{1}{J} \frac{dJ}{dV} = -0.01 J^{-1/2}$,
$\sqrt{S} \frac{1}{U} \frac{dU}{dV} = 0.01 J^{-1/2}$. In all three plots, the
straight dashed green lines correspond to the result obtained in perturbation
theory (see text).}%
\label{fig:cf}%
\end{figure*}
\subsubsection{Numerical results}

In Fig.~\ref{Fig:heating_rates_comparison} we show results for the short time
mean heating rate per particle for unit filling ($\bar n=1$) at different
values for the interactions. The analytical results for weak and strong
interaction are shown together with the results from numerical methods
outlined in Sec. \ref{sec:Methods}. The heating rates obtained with t-DMRG in
one dimension (up to $N=M=30$ particles per lattice site) agree very well with the
analytical results in both limiting cases and connect these smoothly across
the phase transition.

On a mean-field level the (ground state) phase transition from a SF
($|\psi_{l}|^{2}>0$) to an MI ($\psi_{l}=0$) ground state occurs at
$u_{c}\approx5.8z$. The Gutzwiller wavefunction captures the two limiting
cases of an ideal superfluid as a product of coherent states at each lattice
site for vanishing interaction and a Mott insulator as a product of Fock
states at each lattice site. However, a general limitation of the Gutzwiller
mean-field theory is that the entire Mott insulating phase (at integer filling
$\bar n$) is represented by the same wavefunction $|\Psi\rangle=\prod_{l}%
|\bar{n}\rangle_{l}$. This state is trivially invariant under the evolution
with the stochastic equations~\eqref{eq3.1}. As an (unphysical) artifact of
this limitation Eqs.~\eqref{eq3.1} predict no heating in the entire
Mott phase. The only nontrivial dynamics can be observed on the superfluid
side of the phase transition. On this side, except close to the phase
transition, where the mean-field treatment is expected to fail, the results
are in very good agreement with the analytical results obtained from the
Bogoliubov approximation.

\subsection{Evolution of state characteristics}

\label{sec:ResultsB}

The heating of a general many-body quantum state cannot be fully understood by
a single heating rate, since the system is driven out of thermal equilibrium
in general. In order to further quantify the heating, we analyze how
characteristic correlation functions of the different many-body states are
affected by the noisy lattice.

\subsubsection{Single Particle Density Matrix and Condensate fraction}

For the Bose-Hubbard model, the MI and the SF states are characterized by the
off-diagonal correlations, i.e., the off-diagonal elements of the
single-particle density matrix (SPDM), $\langle b^{\dag}_{i} b^{\dag}_{i+j}
\rangle$. The signature of the SF ground state is off-diagonal long-range
order, i.e., these elements decay to a constant (decay algebraically in one dimension),
whereas they decay exponentially to zero in the MI. Here we
analyze how these characteristics change as a function of time.

Closely related to the SPDM is the condensate fraction, which is defined as
the largest of the eigenvalues $\{\lambda_{i}\}$ of the SPDM: $ \mathcal{F}
\equiv\lambda_{0}/\sum_{i} \lambda_{i}  $. Note that in the case of the
Gutzwiller ansatz, the condensate fraction is simply given by $\mathcal{F}%
=|\psi_{l}|^{2}=| \langle\phi_{l}| b_{l} |\phi_{l}\rangle|^{2}$. The
perturbative analysis Sec.~\ref{sec:ResultsA} shows that the main effect of
noise in the lattice depth on a Bose-Einstein condensate is the generation of
quasiparticle pairs, and therefore a decrease of the number of atoms in the
condensate mode with time. As for the heating rate one can calculate this
depletion rate from the master equation \eqref{eqMaster}. Denoting the number of
atoms in the mode with quasi-momentum $q$ by $N_{q}\equiv\langle b_{q}^{\dag
}b_{q} \rangle$ we have $\langle\!\langle\dot{     N}_{q}\rangle
\!\rangle=\frac{S_{0}}{2}\xi\frac{1}{U}\frac{dU}{dV}\langle\left[ \left[
H_{U},b_{q}^{\dag}b_{q}\right] ,H_{U}\right]  \rangle$ for the system
initially in the ground state. In the limit of an ideal condensate the number
of atoms in the condensate mode at $q=0$ evaluates to $\langle
\!\langle\dot{     N}_{0}\rangle\!\rangle/N_{0}=-S_{0}\xi\frac{1}{U}\frac
{dU}{dV}U^{2}\frac{N_{0}}{M}$. Within the same approximation the
quasimomentum of the particles scattered out of the condensate mode is
distributed homogeneously over the whole Brillouin zone: $\langle\!\langle\dot
N_{q}\rangle\!\rangle=S_{0}\xi\frac{1}{U}\frac{dU}{dV}U^{2}\frac{N_{0}^{2}%
}{M^{2}}$.

\begin{figure}[tb]
\centering
\includegraphics[width=0.4\textwidth]{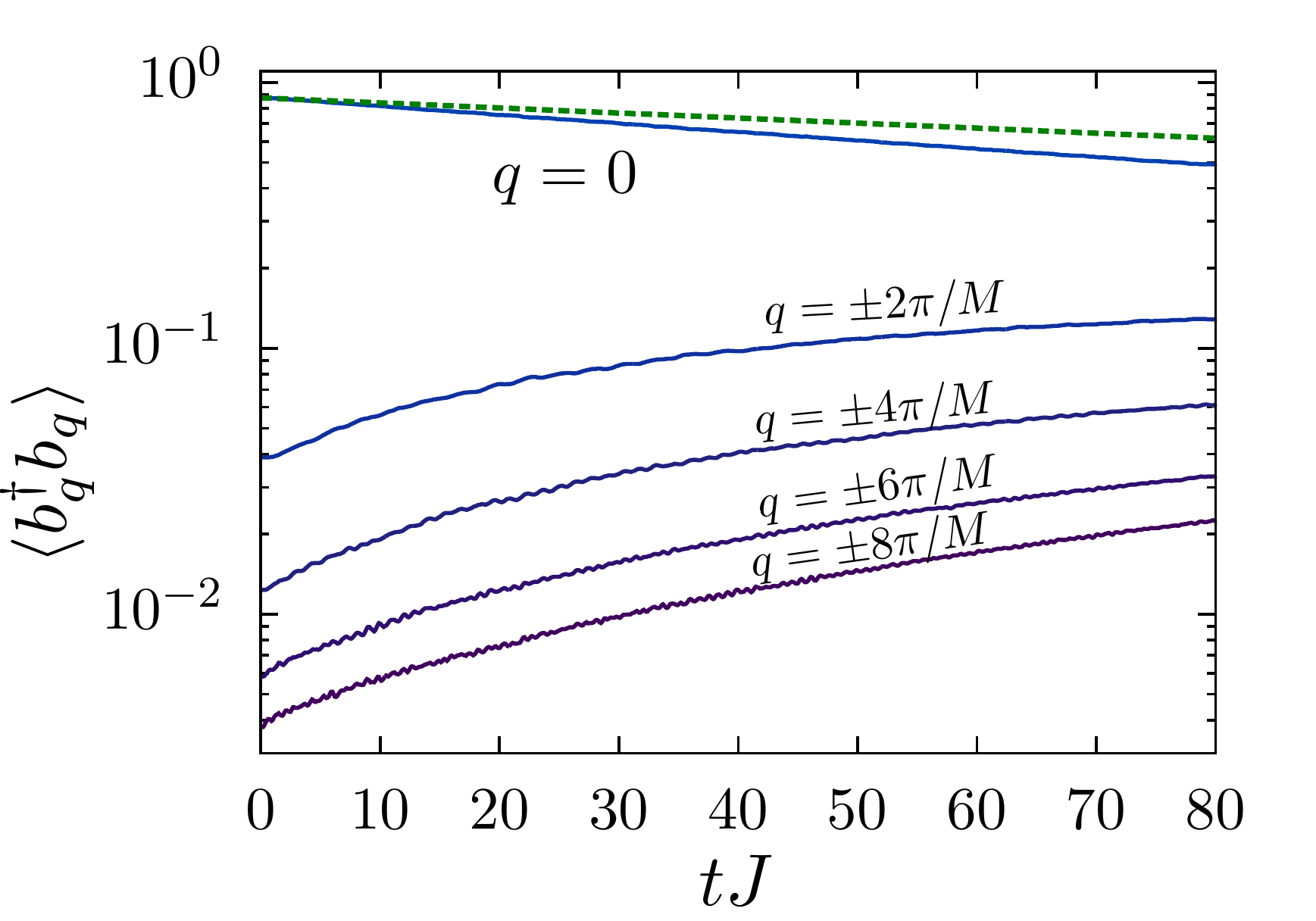}\caption{(Colour online) The evolution of
the population in different quasi momentum modes. Shown are averages over 500
noise realizations for a superfluid state ($U/J=2$) for a 1D system with
$N=10$ particles on $M=10$ sites. The lattice fluctuations lead to a decrease of the
population in the condensate mode at $q=0$ in agreement with the prediction from perturbation theory given by the dashed line. The noise is anticorrelated with $\sqrt{S}
\frac{1}{J} \frac{dJ}{dV} = -0.025 J^{-1/2}$, $\sqrt{S} \frac{1}{U} \frac
{dU}{dV} = 0.025 J^{-1/2}$.}%
\label{fig:mdf}%
\end{figure}

Numerical solutions of the stochastic differential equation show such a
depletion of the condensate mode, both using exact methods [Figs.~\ref{fig:cf}(a),~\ref{fig:cf}(b) and Fig.~\ref{fig:mdf}] (exact diagonalization and t-DMRG) as well as in
the Gutzwiller framework [Fig.~\ref{fig:cf}(c)]. By performing a small noise
expansion \cite{Gar_book} of the nonlinear stochastic Gutzwiller equations we
find that this initial decay of the condensate fraction in the Gutzwiller
framework is associated with the incoherent excitation of the amplitude mode
at zero quasimomentum \cite{Kru11} [see Fig.~\ref{fig:cf}(c)]. However, we
note that for long times solutions to the Gutzwiller equations \eqref{eq3.1}
have a mean condensate fraction of approximately $\sim0.3$. This steady-state
mean condensate fraction is not associated with a steady-state condensate
fraction in the individual trajectory, where in the long-time limit the
condensate fraction oscillates between 0 and 1 with a random phase, which
gives rise to a nonzero steady-state behavior for the stochastic average. Such
a signature is considered an unphysical artifact of the incapacity of the
Gutzwiller wavefunction to capture the decay of the amplitude mode. In fact,
exact simulations of one-dimensional systems show no such behavior.

In Fig.~\ref{fig:cf}(b), we show the stochastic average of the condensate
fraction for a one dimensional system of ten sites and periodic boundary conditions using
exact diagonalization. Even though in one dimension the ground state shows
only quasi-long-range order, for small values of $U/J$ the decay of the
condensate fraction is well described by the expression obtained in
perturbation theory. Our t-DMRG simulations essentially lead to the same
results but are limited to very short times [Fig.~\ref{fig:cf}(a)]. In
contrast to the Gutzwiller results, the condensate fraction decreases
monotonically also for long times for all superfluid initial states.

\subsubsection{Particle-hole correlations}

\begin{figure}[tb]
\centering
\includegraphics[width=0.46\textwidth]{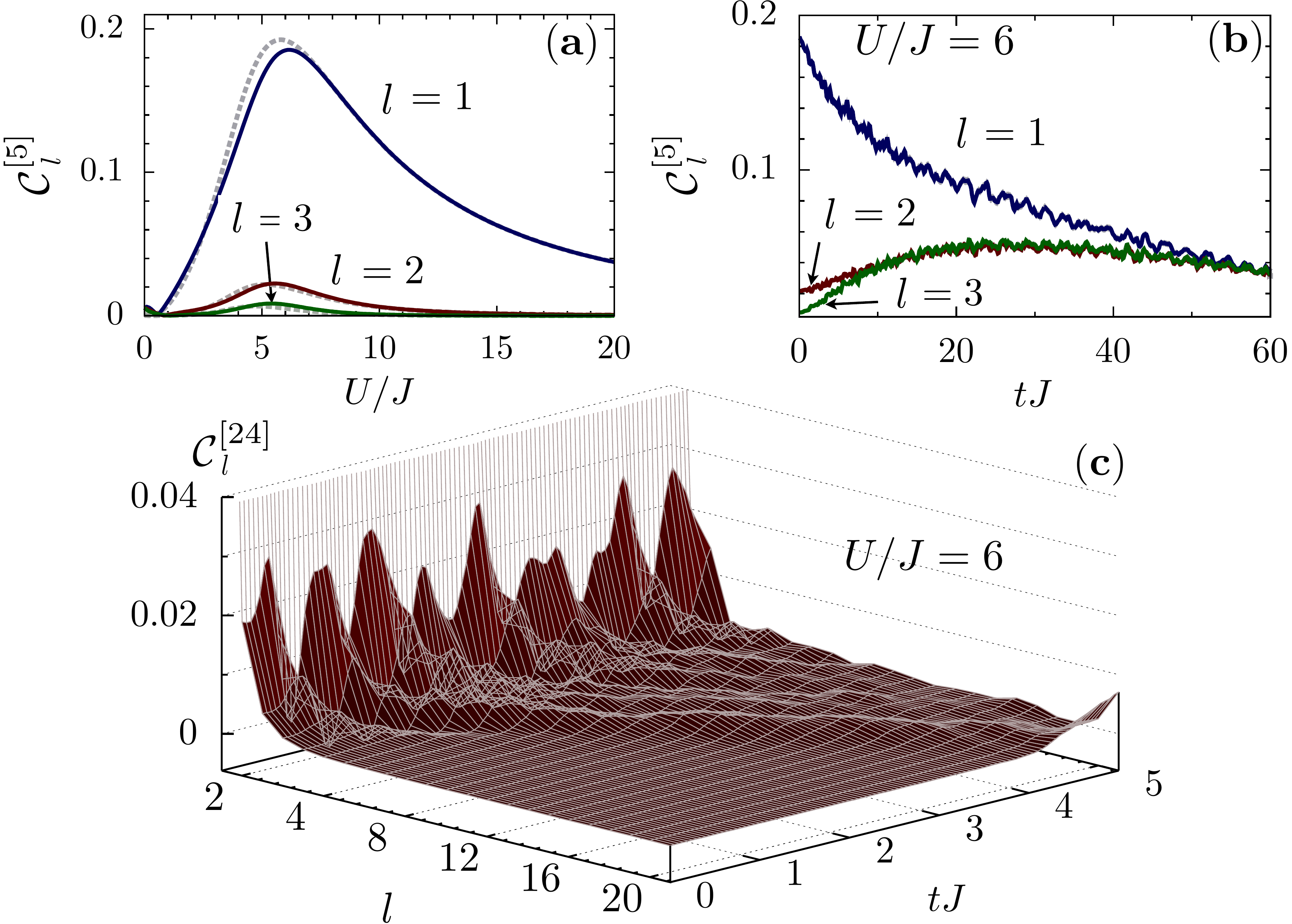}\caption{(Color online) (a)
Parity correlations in the ground state of a small system with $N=M=10$. The
next-neighbor and longer-ranged correlations assume a maximum at $U/J\approx6$
(solid lines are for open, dashed lines are for periodic boundary conditions).
(b) Long-time evolution of these correlations calculated with exact
diagonalization in the small system, averaged over $500$ noise trajectories
(periodic boundary conditions). The transient state develops long-range
correlations. (c) The evolution of the parity-parity correlation function of a
single noise trajectory in a Mott insulating state with $U/J=6$ in a 1D system
with $N=M=48$ (bond dimension $D=265$, local dimension truncation $d_{l}=6$).
The amplitude noise excites single particle-hole pairs which spread out
through the system as a light cone. In all simulations, the noise is
anticorrelated with $\sqrt{S} \frac{1}{J} \frac{dJ}{dV} = -0.025 J^{-1/2}$,
$\sqrt{S} \frac{1}{U} \frac{dU}{dV} = 0.025 J^{-1/2}$ .}%
\label{fig:dubl_hol_corr}%
\end{figure}

In the limit of strong interactions, as shown above, we expect elementary
excitations to consist of correlated particle-hole pairs propagating through
the system. To analyze the dynamics associated with these excitations, we
compute parity correlation functions defined as \cite{Che12}
\begin{align}
\label{eq:parity-correlation}C^{[i]}_{l} = \langle s_{i} s_{i+l} \rangle-
\langle s_{i} \rangle\langle s_{i+l} \rangle,
\end{align}
where $s_{j}$ is the local parity operator at site $j$, defined as $s_{j}%
=\exp{ \{ i \pi(n_{j} - \bar n) \} }$. Since we calculate these functions in
finite inhomogeneous 1D systems, $C^{[i]}_{l}$ depends on the site $i$ from
which we start calculating the function, and we will typically begin from the
central site $i=M/2$ in a system of $M$ sites. In Fig.~\ref{fig:dubl_hol_corr}
we plot the evolution of this correlation function under the evolution with
the SMBSE, calculated with t-DMRG for a large system and with exact
diagonalization for a small system. Note that useful information about these
correlations cannot be obtained from a Gutzwiller product state ansatz, as the
first term of~\eqref{eq:parity-correlation} will always factorize and thus the
correlation will always be zero.

As seen in Fig.~\ref{fig:dubl_hol_corr}(a), the parity correlations in the
ground state assume the largest value in the Mott insulating phase at
$U/J\approx6$. In a noisy time evolution we find that the initially large
next-neighbor parity correlation starts to decrease, whereas the long-range
parity correlations start to increase, as shown in
Fig.~\ref{fig:dubl_hol_corr}(b). It reaches a maximum at a transient state at
times $tJ\approx30$, where the next-neighbor and long-range interactions
assume nearly the same value. At longer times all these correlations start to
decrease into a more and more classical state without correlations.
Furthermore, looking at a single noise trajectory in a large system, we find
in Fig.~\ref{fig:dubl_hol_corr}(c) that the elementary excitations induced by
the amplitude noise are seen as excitations in the parity-parity correlations,
which spread out in the form of a light cone, similar to the results in
\cite{Che12}, where these correlation functions are directly measured in a
quantum gas microscope experiment.

\section{Summary and Outlook}

\label{sec:Summary}

In conclusion we have derived a microscopic model for the dynamics of bosonic
atoms in a time-dependent optical lattice. 
While controlled periodic modulation of the lattice has the potential to access interesting physics \cite{Gre11,Tok11}, we considered the alternative situation where the lattice depth fluctuates stochastically 
at low frequency. This situation arises naturally in optical lattice experiments due to intensity fluctuations of the lattice lasers. Using analytical approximations as well as t-DMRG and
time-dependent Gutzwiller methods, we have analyzed the nonequilibrium
dynamics of many-body states in a variety of parameter regimes. We find
characteristic responses of the system that vary in different parameter
regimes, and could be used to identify the effects of such dynamics in current experiments.

The generalization of the initial model to fermionic atoms results in a
stochastic equation of motion similar to \eqref{eqStratonovich},
\begin{align}
(S)\quad & d|\Psi\rangle=-iH|\Psi\rangle dt-iH^{\prime}\sqrt{S_{0}}%
|\Psi\rangle dW_{t},
\end{align}
with a coherent part given by the Fermi-Hubbard Hamiltonian, and a
corresponding stochastic contribution:
\begin{align}
H &  = -J \sum_{\langle i,j\rangle,\sigma}c_{i,\sigma}^{\dag}c_{j,\sigma}
+U\sum_{i} c_{i,\uparrow}^{\dag}c_{i,\uparrow}c_{i,\downarrow}^{\dag
}c_{i,\downarrow},\\
H^{\prime} &  = -\frac{dJ}{dV} \sum_{\langle i,j\rangle,\sigma}c_{i,\sigma
}^{\dag}c_{j,\sigma} +\frac{dU}{dV}\sum_{i} c_{i,\uparrow}^{\dag}%
c_{i,\uparrow}c_{i,\downarrow}^{\dag}c_{i,\downarrow}.
\end{align}
Here the $c_{i,\sigma}$ are operators annihilating a fermion at site $i$ with
spin $\sigma\in\{\uparrow,\downarrow\}$. An analysis of this equation, similar
to the one presented in Sec.~\ref{sec:Results}, can be carried out. In the Mott
insulator at half filling for the antiferromagnetic ground state ($U\gg J$) a
heating rate of $\dot E /N=S_{0} \xi^{2} U zJ^{2}$ can be found from
mean-field calculations. Exploration of heating in other regimes, e.g.,\ in the
BCS-BEC crossover, is an interesting direction for further analysis.

\section*{Acknowledgments}

We thank P.~Barmettler, S.~Blatt, M.~Lukin, C.~Spee, and the groups of I.~Bloch and
W.~Ketterle for helpful and motivating discussions. H.P. thanks the University
of Pittsburgh for hospitality. This work was supported in part by the Austrian
Science Fund through SFB F40 FOQUS, and by a grant from the U.S. Army Research
Office with funding from the DARPA OLE program. Work in Pittsburgh was
supported by AFOSR Grant No. FA9550-12-1-0057, and computational resources were
provided by the Center for Simulation and Modeling at the University of
Pittsburgh.  \newline\appendix

\section{Interband transitions}\label{app:Interband}
Here we comment on the derivation of the SMBSE focusing on the transitions to higher bands due to nonadiabatic transitions.

To proceed from Eq.~\eqref{eq2.6} we expand the field operator in the instantaneous Wannier basis. The coefficients of the wavefunction $\ket{\Psi}$ in the corresponding instantaneous Fock basis $\ket{\{n_{i,n}\}}$ change in time via the change of the wavefunction and via the change of the time-dependent basis states:
\begin{align}
\frac{d}{dt}\braket{\{n_{i,n}\}}{\Psi}=\lr{\frac{d}{dt}\bra{\{n_{i,n}\}}}\ket{\Psi}+\bra{\{n_{i,n}\}}\lr{\frac{d}{dt}\ket{\Psi}}.
\end{align}
The time derivative of a Fock basis state $\ket{\{n_{i,n}\}}$ constructed from the time dependent single particle basis
$
w_{i,n}(x,V(t))$ can be obtained from the time derivative of the
corresponding annihilation operators: $b_{i,n}=\int dx w_{i,n}(x,V(t))\hat\psi(x)$. With $\psi(x)=\sum_{i,n}w_{i,n}(x,V(t)) b_{i,n}$ we
find
\begin{align}
\frac{d}{dt}b_{i,n}  & =\sum_{i,n}\int dx\left(  \frac{d}{dt}w_{i,n}(x,V(t)) \right)  w_{i,n}(x,V(t)) b_{i,n}.
\end{align}
This expresses the change of the annihilation operators in the Schr\"odinger
picture due to the change of the basis states. It should not be confused with
the Heisenberg equation of motion for this operators. With this relation it is
easy to show that the corresponding Fock states $|\{n_{i,n}\}\rangle$
change due to the changing single particle basis as
\begin{align}
&\frac{d}{dt}|\{n_{i,n}\}\rangle =\no\\
&\sum_{r,s,j,m}\int dx\, w_{j,m}(x,V(t))\left(\frac{d}{dt}w_{r,s}(x,V(t)) \right)  b_{j,m}^{\dag}b_{r,s}\ket{\{n_{i,n}\}},
\end{align}
which gives rise to the term $G(V(t))$ in Eq.~\eqref{eq2.9}. A similar derivation is given in
\cite{Lac12}.
To see that this term couples only states in different bands it is instructive to transform $G(V)$ to the instantaneous Bloch basis
 $\phi_{q,n}(x,V)$ (with the
corresponding annihilation operator $b_{q,n}$), 
\begin{align}
G(V)=i\!\!\!\sum_{q,n;p,m}\!\int dx\, \phi_{p,m}(x,V)\frac{d \phi_{q,n}^{\ast
}(x,V)}{dV} b_{q,n}^{\dag}b_{p,m}.\label{eq2.12}%
\end{align}
The derivative of the Bloch state $\phi_{q,n}(x,V)$ with respect to the
lattice depth can be expressed in terms of the Bloch states with the same
quasimomentum $q$, but in different bands $m\neq n$. From first-order
perturbation theory we find
\begin{align}
& \frac{d}{dV}\phi_{q,n}(x,V)=\sum_{m\neq n} \phi_{q,m}(x,V)\sigma
_{q,m}(V),\label{eq2.13}\\
& \sigma_{q,m}(V)=\frac{\int dy \phi^{\ast}_{q,m}(y,V)\phi_{q,n}(y,V) \sin
^{2}(ky)}{\varepsilon_{q,n}(V)-\varepsilon_{q,m}(V)}.\label{eq2.14}%
\end{align}
Thus the integral in~\eqref{eq2.12} is nonzero only for $p=q$ and for $n\neq
m$. The operator $G(V)$ therefore transfers single atoms into a different band
with the same quasimomentum. This conservation of quasimomentum reflects the
fact that the lattice translation symmetry is not broken by global
fluctuations in the lattice depth. Further, the perturbation does not break
the reflection symmetry $x\rightarrow-x$. As a consequence, bands with even
(odd) band index $n$ are coupled only to bands with an even (odd) band index
$m$.

To lowest order in $J/|\omega_{n,m}|$, with $\omega_{n,m}$ being the energy difference between the bands $n$ and $m$, we can find a simple expression for the
resonant parts in $G(V_{0})\delta\dot V(t)$ that affect atoms in the lowest
band:
\begin{align}
\label{eq2.15} G(V_{0})\delta\dot V(t)&\approx\sum_{i,n> 0}\lambda
_{i,n}\delta V_{n}(t)b_{i,n}^{\dag}b_{i,0}+\text{H.c.},\\
\lambda_{i,n}&=\int dx\, w_{i,n}(x)\sin^{2}(kx)w_{i,0}(x),
\end{align} where we
again kept only terms diagonal in the site index $i$ due to the localized form
of the Wannier functions. Here $\delta V_n(t)$ denotes restriction of the noise term $\delta V(t)$ to frequencies around transition frequency from the lowest to the $n$th band. Approximating the Wannier functions at a given
lattice site with harmonic oscillator wavefunctions, we find that atoms in the
lowest band are dominantly scattered into the second excited band in such
band-changing processes. To lowest order in the Lamb Dicke parameter
$\eta=ka_{0}$, which measures the extension of the Wannier function $a_{0}$,
compared to the lattice constant $a=\pi/k$, one finds $\lambda_{i,n}%
\approx\sqrt{2}\eta^{2}\delta_{n,2}+\mathcal{O}(\eta^{4})$. Thus, the number of
atoms in the lowest band $N_{0}$ decreases in time as $\dot N_{0}/N_{0}%
=-\sum_{n>0}S_{n}\lambda_{i,n}^{2}\approx-2\eta^{4} S_{2}$. This sets the
timescale on which the restriction to the lowest band is valid.

\end{document}